# The rest mass of a system of two photons in different inertial reference frames: 0+0=0 and 0+0>0

**Bernhard Rothenstein**, Physics Department,
"Politehnica" University of Timişoara, Romania

**Doru Păunescu**, Department of Mathematics,
"Politehnica" University of Timişoara, Romania

**Ştefan Popescu**, Siemens, Erlangen, Germany

**Abstract.** *We show that the rest mass of a system consisting of two photons is a relativistic invariant having the same magnitude in all inertial reference frames in relative motion. A scenario which starts with two photons where theirs frequencies are equal to each other the magnitude of the rest mass of the system depends on the angle $\theta'$, made by the momentums of the two photons being equal to zero for $\theta'=0$ and $\theta'=2\pi$ and maximum $(2h\nu')$ for $\theta'=\pi$. The behavior of the photons when detected from two inertial reference frames is illustrated using a relativistic diagram which displays in true values the physical quantities (scalar and vector) introduced in order to characterize the dynamical properties of the photon. We consider that the obtained results should be taken in the discussions, some time fierce, between those who ban the concept of relativistic mass and those who consider that its use is harmless.*

1. ## Introduction

Physicists operate with two kinds of particles: **tardyons** and **photons (luxons).** A tardyon is a particle the speed of which never exceeds the speed $c$ at which electromagnetic radiation propagates through empty space. The electron is an outstanding member of that class of particles. The photon is a particle that exists only moving with speed $c$. In accordance with the second postulate of Einstein's special relativity the photon moves with $c$ relative to all inertial observers in relative motion. Existing only in a state of motion relativists consider that the **rest mass** of a photon is equal to zero.

Tackling the problem in the limits of special relativity theory we involve the observers $R$ of the $K(XOY)$ reference frame and the observers $R'$ of the $K'(X'O'Y')$ inertial reference frame, both inertial ones. The corresponding axes of the two frames are parallel to each other the $OX(O'X')$ axes being overlapped. At the origin of time in the two frames $(t=t'=0)$ the origins of the two frames are located at the same point in space, $K'$ moving with constant speed $V$ relative to $K$ in the positive direction of the overlapped axes. Physical quantities measured in $K$ are presented as unprimed whereas physical quantities measured in $K'$ are presented as primed. We do not make that distinction in the case of $c$.

A photon located at $t'=0$ at the origin $O'$ of $K'$ generates after a time $t'$ of motion the event $E'(x'=r'\cos\theta', y'=r'\sin\theta', t'=r'/c)$ using polar $(r',\theta')$ and Cartesian $(x',y')$ space coordinates in a two space dimensions approach. Detected from $K$ the **same** event is characterized by the space-time coordinates $E(x=r\cos\theta, y=r\sin\theta, t=r/c)$. The Lorentz-Einstein transformations for the space-time coordinates of the same event[1] tell us that

$$x = \gamma r'(\cos\theta' + \beta) \qquad (1)$$
$$y = r'\sin\theta' \qquad (2)$$
$$r = \gamma r'(1+\beta\cos\theta') \qquad (3)$$
$$t = \gamma t'(1+\beta\cos\theta') \qquad (4)$$
$$\cos\theta = \frac{\cos\theta' + \beta}{1+\beta\cos\theta'} \qquad (5)$$

with the usual relativistic notations $\beta = V/c$ and $\gamma = (1-\beta^2)^{-1/2}$.

Classical electrodynamics tells us[2], in accordance with experimental facts thatr electromagnetic radiation carries energy and momentum because it heats and exerts pressure on the objects on which it is incident. Let $E_{e.m.}$ and $p_{e.m.}$ be the energy and the momentum carried by the electromagnetic radiation respectively related by

$$p_{e.m.} = \frac{E_{e.m.}}{c} \qquad (6)$$

The energy and momentum being carried by an invariant number of photons we obtain that momentum and energy of a photon are related by

$$p = \frac{E}{c}. \qquad (7)$$

Even opponents of the concept of **relativistic mass** agree with the fact that a photon has a relativistic mass

$$m = \frac{p}{c} = \frac{E}{c^2}. \qquad (8)$$

The **same** photon, detected from $K'$ is characterized by the momentum $p'$, energy $E'$ and relativistic mass $m'$ related by

$$p' = \frac{E'}{c} = cm'. \qquad (9)$$

In order to explain the outcome of the interaction between an incident photon and a conduction electron (photoelectric effect) Einstein considered that the energy, momentum and relativistic mass of a photon and the frequency of the electromagnetic oscillations taking place in the associated electromagnetic wave $\nu$ are related by

$$E = cp = c^2 m = h\nu \qquad (10)$$

in $K$ and by

$$E' = cp' = c^2 m' = h\nu' \qquad (11)$$

$h$ representing Planck's constant.
Equations (10) and (11) suggest that the transformation equations for the energy, momentum and relativistic mass of the same photon, goes through the **Doppler shift formula**[3] that establishes the following relationship between the frequencies $\nu$ and $\nu'$

$$v = \gamma v'(1+\beta\cos\theta') \qquad (12)$$

and so

$$E = \gamma E'(1+\beta\cos\theta') \qquad (13)$$
$$p = \gamma p'(1+\beta\cos\theta') \qquad (14)$$
$$m = \gamma m'(1+\beta\cos\theta'). \qquad (15)$$

Taking into account the vector character of the momentum, the components of the momentum transform as

$$p_x = \gamma p'(\cos\theta' + \beta) \qquad (16)$$
$$p_y = p'_y. \qquad (17)$$

In the case of a tardyon with rest mass $m_0$, having energy $E_t$ and momentum $p_t$ when detected from $K$ we have[4]

$$E_t^2 - c^2 p_t^2 = m_0^2 c^4. \qquad (18)$$

Applied in the case of a photon ($E = cp$) equation (18) leads to

$$E^2 - c^2 p^2 = 0, \qquad (19)$$

confirming the zero value of the rest mass of a photon.

In the discussions, sometimes fierce, between those who ban the concept of relativistic mass (contras) taking into account only the concept of rest mass and those who consider that the use of the concept of relativistic mass is harmless (pros), a serious point is the fact that **the rest mass of a system of two photons is not always equal to zero**[5,6]. The problem is discussed in detail[7] in a one space dimensions approach. and from a single reference frame the frequency of the involved photon being not Doppler shifted.

## 2. A relativistic diagram for illustrating the behavior of a photon when we detect it from different inertial reference frames in relative motion

Let $K'$ be the reference frame where a source of monochromatic radiation $S'(0,0)$ located at its origin emits light signals (photons) in all directions of the plane defined by the axes of the considered reference frame. Observers of that frame attribute to the photons emitted under such conditions the same frequency $v'$. Observers from $K$ attribute to the same photons Doppler shifted frequencies in accordance with (12). The equations derived above suggest considering the diagram[8] presented in Figure 1 on which we have overlapped a circular wave front detected from $K'$

$$r' = ct' \qquad (20)$$

and the **same** wave front detected from $K$[9]

$$r = \gamma r'(1+\beta\cos\theta') = \gamma ct'(1+\beta\cos\theta'). \qquad (21)$$

The circle (20) has its center at the point where the origins $O$ and $O'$ are located at the origin of time in the two frames whereas one of the foci of the ellipse (21) is located at that point as well. Consider that event $E'$ defined above takes place on the wave front detected from $K'$. The same event $E'$ detected from $K$ takes place on the same wave front detected from $K$. The invariance of distances measured normal to the direction of the relative motion of the two reference frames enables us to find out the relative positions of the two events on our relativistic diagram as shown in **Figure 1**. It is easy to show that the diagram displays in true magnitudes the space-time coordinates of the two events in accordance with the values obtained performing theirs Lorentz-Einstein transformations.

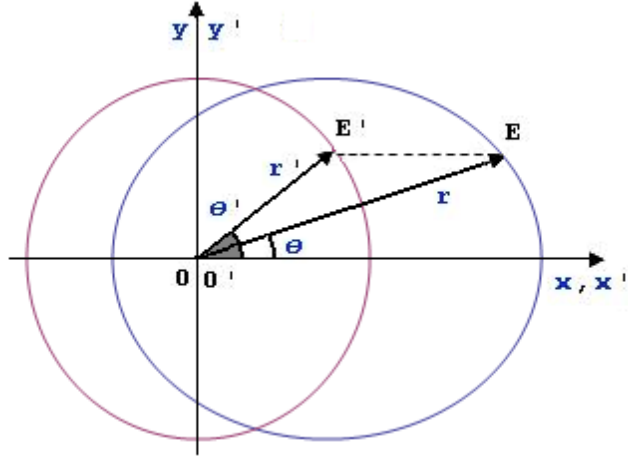

**Figure 1.** *Events E and E', generated by the same photon and detected from the K and K' inertial reference frames on a relativistic diagram which displays in true magnitudes the polar and the Cartesian coordinates of the events.*

Taking into account that the Lorentz-Einstein transformations
$$(x \to x'; p_x \to p'_x); (y \to y'; p_y \to p'_y); (r \to r'; p \to p') \text{ and } (t \to t')$$
take place via the same transformation factors respectively we can construct the relativistic diagram presented in **Figure 2** where the circle accounts for the isotropic emissions of photons in K' its radius equating $E'$, $c^2 m'$, $h\nu'$, $cp'$ where they have the same magnitude in all directions. whereas the ellipse described by (12), (13), (14) and (15) accounts for theirs magnitudes detected from K. The invariance of the *OY(O'Y')* components of the momentum helps us to find out that if in K' the end of the momentum vector $\mathbf{p}'(p'_x, p'_y)$ is located on the circle at point *1'* then detected from K is is located on the ellipse at the point *1*. The diagram displays in true values the physical quantities introduced in order to characterize the same photon as detected from two inertial reference frames in relative motion.

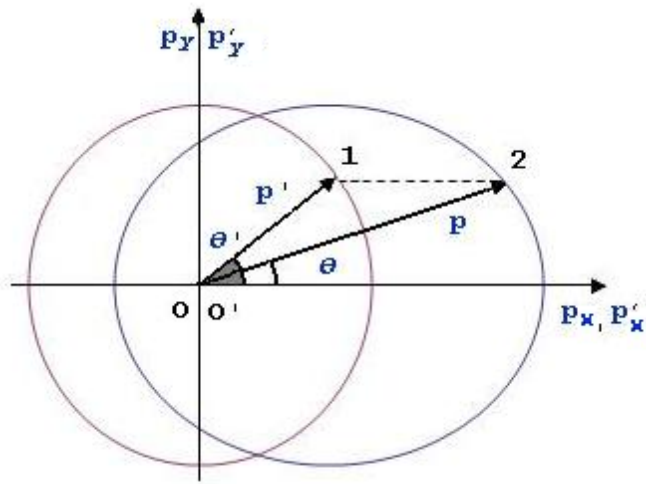

**Figure 2.** *The relativistic diagram that displays in true values the magnitudes of the physical quantities introduced in order to characterize the dynamic properties of a photon as detected from two inertial reference frames in relative motion.*

## 3. The system of two photons detected from two inertial reference frames in relative motion.

Let *K'* be the reference frame where all the considered photons have the same frequency *v'*. Consider a photon emitted in the positive direction of the *O'X'* axis (index 1) and a second photon (index 2) emitted along a direction that makes an angle *θ'* with the positive direction of the *O'X'* axis which could change in the range of values *0<θ'<2π* as shown in **Figure 3**.

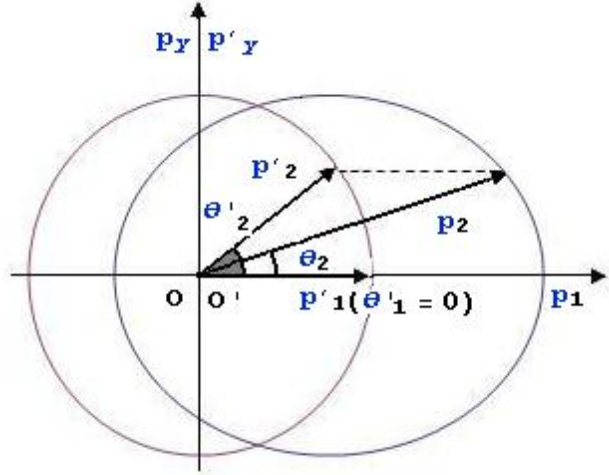

**Figure 3.** *The system of two photons $p'_1$ and $p'_2$ as detected from K' and from K respectively ($p_1$ and $p_2$).*

Equation (18) applied in the case of the system of two photons leads to

$$m_{0,1,2} = \sqrt{\frac{(E_1+E_2)^2}{c^4} - \frac{(\mathbf{p}_1+\mathbf{p}_2)^2}{c^2}}. \tag{22}$$

The energy of the system of two photons is in *K*

$$E_{1,2} = E_1 + E_2 = hv'\left[\sqrt{\frac{1+\beta}{1-\beta}} + \frac{1+\beta\cos\theta'}{\sqrt{1-\beta^2}}\right] \tag{23}$$

whereas its momentum is given by

$$\mathbf{p}_1+\mathbf{p}_2 = p_{1,2} = \frac{hv'}{c}\sqrt{\left[\left(\frac{\cos\theta'+\beta}{\sqrt{1-\beta^2}} + \sqrt{\frac{1+\beta}{1-\beta}}\right)^2 + \sin^2\theta'\right]} \tag{24}$$

with which the rest mass of the system of two particles becomes

$$\frac{c^2 m_{0,1,2}}{hv'} = \sqrt{2(1-\cos\theta')} = 2\sin\frac{\theta'}{2} \tag{25}$$

As we see $\dfrac{c^2 m_{0,1,2}}{h\nu'}$ is β independent telling us that the rest mass of the system of two particles is invariant.

We present in **Figure 4** the variation of the dimensionless physical quantity $\dfrac{c^2 m_{0,1,2}}{h\nu'}$ (25) with the angle θ'. As we see, its magnitude is zero for θ'=0 and θ'=2π and has a maximal value for θ'=π. On the same figure we present the variation with θ' of $E_{1,2}$ and $p_{1,2}$ which are, as we can see, θ' independent.

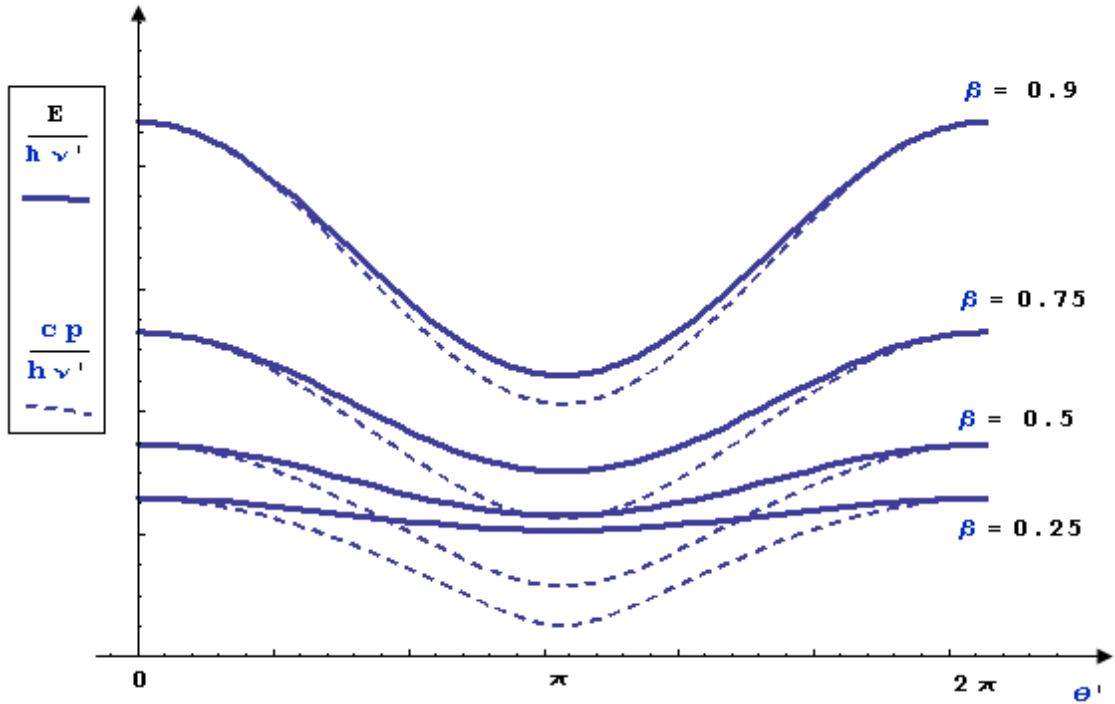

**Figure 4.** *i) The variation of (23) and (24) with θ' illustrating the fact that $E_{1,2}$ and $p_{1,2}$ are β independent.*

We consider that the results presented by us should redirect the discussions between the pros and the contras in the endless debate between the pros and the contras of the concept of relativistic mass.

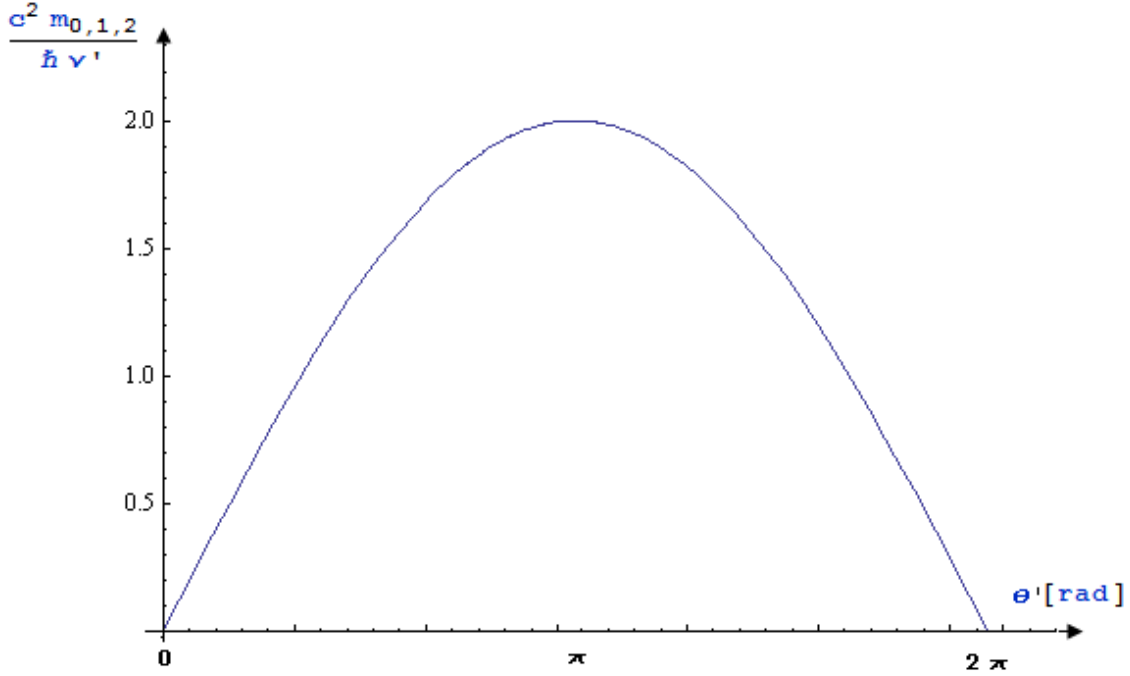

**Figure 4.** *ii) The variation of (25) with θ' illustrating the fact that $m_{0,1,2}$ is β independent.*

## 4. The system of isotropically emitted identical photons

The photon gas is a frequent topic in special relativity.[2] In order to avoid the complications introduced by the vessel in which the photon gas is confined we follow a scenario proposed by Gron[10] considering that a body at rest in *K'* and located at its origin *O'* emits a short pulse of light isotropically all the emitted photons having the same frequency *v'*. Let *N* be the invariant number of emitted photons and *Nhv'* the total emitted energy, theirs resultant momentum being equal to zero. On our relativistic diagram the ends of the momentum of the photons are located on the circle. Detected from *K* they are located on the ellipse.

Consider a pair of photons moving when detected from *K'* along the direction that make the angles θ' and θ'+π with the positive direction of the overlapped axes. In *K*, the energy of the two photons is

$$E_{\theta,\theta+\pi} = hv'[\gamma(1+\beta\cos\theta') + \gamma(1-\beta\cos\theta')] = 2\gamma hv' . \quad (26)$$

The *OY(O'Y')* components of the momentums cancel each other and so the resultant momentum is in *K*

$$p_{\theta,\theta+\pi} = hv'c^{-1}\left(\sqrt{\frac{1+\beta}{1-\beta}} - \sqrt{\frac{1-\beta}{1+\beta}}\right) = 2hc^{-1}\gamma\beta v' \quad (27)$$

both θ' independent. In *K*, the rest mass of the considered system of photons is, in accordance with (18),

$$m_{0,N} = Nhv'c^{-2}\sqrt{\frac{1}{1-\beta^2} - \frac{\beta^2}{1-\beta^2}} = Nhv'c^{-2} \quad (28)$$

being, as we see, a relativistic invariant.

## 5. Conclusions

Consider the problem of the rest mass of a system of two photons in two dimensional space and from two inertial reference frames in relative motion.

By extending the problem of the mass of the system of two photons to two dimensional space, we have brought new insights to it, visualizing the part played by different factors. We have revealed the important fact that its rest mass is a relativistic invariant, having the same magnitude in all inertial reference frames and with its magnitude depending on the angle made by the momentum vectors of the two photons in the reference frame where theirs frequencies are equal.

**E-mail**
Bernard Rothenstein          *Bernard.Rothenstein@mat.upt.ro*
Doru Păunescu                *doru.paunescu@mat.upt.ro*